\documentclass[IOP,onecolumn]{emulateapj}
\newcommand{\myemail}{RK: rony.keppens@wis.kuleuven.be}

\usepackage{graphics}
\usepackage{graphicx}
\usepackage{amsmath,amssymb}

\newcommand{\FIG}[1]{#1}
\newcommand{\PREP}[1]{}

\shorttitle{Tilt-kink repelling current channels}
\shortauthors{Keppens et al.}

\usepackage{tikz}

\begin{document}
%\draft
\title{Interacting tilt and kink instabilities in repelling current channels}
\author{R. Keppens$ ^{1}$, O. Porth$ ^{1,2}$ and C. Xia$ ^{1}$}
%\affiliation{Centre for mathematical Plasma-Astrophysics, Department of Mathematics, KU Leuven, Belgium}
\affil{Centre for mathematical Plasma-Astrophysics, Department of Mathematics, KU Leuven, Belgium \\
Department of Applied Mathematics, The University of Leeds, Leeds LS2 9JT, UK
}
%\author{O. Porth}
%\affiliation{Department of Applied Mathematics, The University of Leeds, Leeds LS2 9JT, UK}
%\author{C. Xia}
%\affiliation{Centre for mathematical Plasma-Astrophysics, Department of Mathematics, KU Leuven, Belgium}
%\maketitle

\email{\myemail}

\begin{abstract}

We present a numerical study in resistive magnetohydrodynamics where the initial equilibrium configuration contains adjacent, oppositely directed, parallel current channels. Since oppositely directed current channels repel, the equilibrium is liable to an ideal magnetohydrodynamic tilt instability. This tilt evolution, previously studied in planar settings, involves two magnetic islands or fluxropes, which on Alfv\'enic timescales undergo a combined rotation and separation. This in turn leads to the creation of (near) singular current layers, posing severe challenges to numerical approaches. Using our open-source grid-adaptive MPI-AMRVAC software, we revisit the planar evolution case in compressible MHD, as well as its extension to 2.5D and full 3D scenarios. As long as the third dimension remains ignorable, pure tilt evolutions result which are hardly affected by out of plane magnetic field components. In all 2.5D runs, our simulations do show secondary tearing type disruptions throughout the near singular current sheets in the far nonlinear saturation regime. In full 3D, both current channels can be liable to additional ideal kink deformations. We discuss the effects of having both tilt and kink instabilities acting simultaneously in the violent, reconnection dominated evolution. In 3D, both the tilt and the kink instabilities can be stabilized by tension forces. As a concrete space plasma application, we argue that interacting tilt-kink instabilities in repelling current channels provide a novel route to initiate solar coronal mass ejections, distinctly different from currently favored pure kink or torus instability routes.
\end{abstract}

\keywords{instabilities -- MHD -- Sun: coronal mass ejections (CMEs) -- Sun: flares}

\section{Introduction}\label{s-intro}

Ideal magnetohydrodynamic (MHD) instability routes represent the most common disruptive scenarios to magnetically dominated plasmas, to which both fusion oriented, laboratory plasma configurations and the intricate magnetic loop systems in the solar corona belong. In laboratory settings, linear MHD stability criteria relevant for ideal MHD, force-balanced states, have introduced operational limits to achievable plasma beta (i.e. the ratio of plasma to magnetic pressure) ranges, with perhaps the most famous Kruskal-Shafranov limit on the total plasma current set from external kink mode stability~\citep[see, e.g.][]{book1}. This instability redistributes poloidal field in a cylindrical (and in a toroidal) plasma-vacuum setup. In a cylindrical $(r,\phi,z)$ single plasma column, it relates to magnetic pressure causing runaway displacement in a $z$-pinch when kink deformed (i.e. $m=1$ for a $\exp(i\,m\phi+k_z\,z)$ perturbation). Known means to stabilize plasma columns against these instabilities involve wall stabilization, and/or eliminating their potential to fit into the column (or torus), with the latter giving the Kruskal-Shafranov result. Although the overall MHD stability of a cylindrical plasma configuration is in reality highly dependent on the actual equilibrium profiles (as function of radius of the cylinder) with internal modes and resonant surfaces causing significant additional complications~\citep{book1,book2}, kink instability is to be avoided for stable laboratory plasma operation, and is also known to relate to violent eruptions in solar coronal loop systems, where the radial twist profile in a fluxrope is key. Fluxropes are found throughout the solar corona, and in essence overlay and connect bipolar flux concentrations on the solar photosphere. If the twist internal to the fluxrope exceeds a critical value (in line-tied, force-free loops studied by~\cite{hood81}), the resulting helical ideal kink evolution can explain both confined (or failed) as well as fully ejective eruption scenarios, as demonstrated by~\cite{torok05}. 

In the solar coronal context, the kink instability is but one of several routes to initiate coronal mass ejections. Again borrowing on laboratory plasma theory~\citep{bateman78}, the torus instability~\citep{kliem06} relies on the fact that a current ring, with major radius $R$ from the ring center, is susceptible to a runaway $R$-directed expansion if an external poloidal field has insufficient stabilizing influence, i.e. if its $R$-variation decreases sufficiently fast. Note that this torus instability does not involve the details of the field variation internal to the (assumed slender) ring, and can not be stabilized by toroidal fields in the fluxrope. Recent observational analysis has confirmed that some fluxrope evolutions could be triggered by torus unstable setups~\citep{zuccarello14}, by using combined Solar Dynamics Observatory (SDO) and STEREO (Solar TErrestrial RElations Observatory) data to infer the three-dimensional magnetic topology of the ambient field conditions in a filament eruption.

Contrary to both kink and torus instability routes for coronal fluxropes, we here present an as yet underexplored route to violent plasma disruptions, where both kink and so-called tilt instabilities~\citep{richard90} interplay. While in both scenarios discussed above, only a single fluxrope or current channel enters the description, the tilt instability relates to the basic fact that two adjacent anti-parallel current channels want to repel each other. Contrary to the torus setup, it does not require toroidal curvature of the fluxropes, as we will study evolutions starting from two adjacent straight current channels, carrying oppositely directed currents. The tilt instability is intimately connected to the coalescence instability, which instead attracts and merges like-directed current channels. Both tilt and coalescence instability have been studied extensively in pure 2D, planar configurations~\citep[see, e.g.][]{longcope93,strauss98,marliani99,ng08}, where the poloidal field distribution forms islands that repel or attract, causing localized reconnection and strong hints for singular current layer development. The singular nature of the current concentrations has turned them into popular testbeds for adaptive mesh refinement (AMR) strategies in MHD simulations~\citep{strauss98,lankalapalli07,ng08}. We here use high-resolution, fixed grid as wel as AMR simulations to study how adjacent repelling current channels, without initial curvature or line-tying, evolve through combined tilt and kink evolutions, in up to 3D configurations.

It is to be noted that the tilt instability is commonly studied in true 3D toroidal setups from laboratory plasmas~\citep[see, e.g.][]{belova06,macnab07}, like the spheromak or field-reversed configuration (FRC, a compact toroid with negligible toroidal field, with poloidal field confinement due to a toroidal plasma current). In the context of FRC and spheromak configurations, the tilt instability becomes the most dangerous global mode causing disruptions.  
Like its pure 2D variant, it tilts the now toroidally symmetric (i.e. $\phi$-direction has $\partial \phi=0$) adjacent (at the symmetry axis $z=0$) cross-sections in a $m=1$ fashion, although the precise shape and elongation of the poloidal flux surfaces can influence the growthrate~\citep{iwasawa00,iwasawapop00}. Recently, the tilt instability of two interacting spheromaks leading to reconnection at the central moving magnetic null point was simulated as well~\citep{lukin11}, as a route to fast reconnection. In contrast to all these essentially toroidal setups, we here deliberately focus on parallel, straight current channels or fluxropes. Our model is representative for the top parts of adjacent loop systems as seen in many extreme ultraviolet views of the highly structured solar corona. Whenever two adjacent loops develop antiparallel currents, the tilt instability route may be accessible. We will show that both internal and external field variations matter, and especially the presence of a toroidal (loop-aligned) component can have a stabilizing effect. More importantly, when instability sets in, synthetic extreme ultraviolet (EUV) views of the optically thin plasma emission show tell-tale signatures of interacting current channels, which must have clear observational counterparts.

The paper is organized as follows. In Section~\ref{s-setup}, we provide details on the actual MHD equilibrium configuration studied, and on the numerical approach. In Section~\ref{s-25d}, we discuss the 2.5D simulations, that confirm and extend previous studies on pure tilt evolutions. In Section~\ref{s-3d}, we discuss our 3D simulations, showing the novel interaction routes between tilt and kink deformations over a fair range of prevailing plasma beta. For the solar application, synthetic EUV views and interpretations are provided in the closing Section~\ref{s-disc}.

\section{Numerical setup}\label{s-setup}

Cartesian $(x,y)$ coordinates are used on a square region $[-3,3]\times [-3,3]$, with vector $z$-components orthogonal to this plane in both 2.5D and 3D cases. The initial condition is determined as follows. Using polar coordinates in the $(x,y)$ plane from $(r,\theta)\equiv\left(\sqrt{x^2+y^2},\arctan(y/x)\right)$, we introduce a flux function $\psi_0(x,y)$ as
\begin{eqnarray}
\psi_0(x,y) & = & \begin{cases} \frac{2}{j^1_0 J_0(j^1_0)} J_1(j^1_0\, r) \cos(\theta)  & \text{for $r< 1$}, \\ 
                                \left(r -\frac{1}{r}\right) \cos(\theta) & \text{for $r\geq 1$} \,. \end{cases} 
\label{q-psi0}
\end{eqnarray}
In this expression, $j^1_0\approx 3.831706$ denotes the first root of $J_1$, and Bessel functions of the first kind are written as $J_n$. We then deduce the magnetic field components as
\begin{eqnarray}
B_x & = & +\frac{\partial \psi_0}{\partial y} \,, \nonumber \\
B_y & = & -\frac{\partial \psi_0}{\partial x} \,, \nonumber \\
B_z & = & B_{z0} \,, 
\label{q-bval}
\end{eqnarray}
where we introduce an additional parameter $B_{z0}$ indicating the strength of the vertical magnetic field component. The resulting current distribution is purely contained within the unit circle, and has two anti-parallel ($j_z=\left(\nabla\times \mathbf{B}\right)_z >0$ and $j_z<0$) current channels that each occupy half the unit disk initially. An ideal MHD equilibrium balance between pressure gradient and Lorentz force is established when setting
\begin{eqnarray}
p(x,y) & = & \begin{cases} p_0+ \frac{(j^1_0)^2}{2} (\psi_0(x,y))^2  & \text{for $r< 1$}, \\
                           p_0  & \text{for $r\geq 1$} \,. \end{cases}
\label{q-pval}
\end{eqnarray}
Note that the original work by~\citet{richard90} typically used a force-free magnetic field using a spatially varying vertical magnetic field component with a uniform plasma pressure, although runs with pressure gradient variations were stated to yield similar, though slightly `slower' evolutions. As we will confirm later on, the configuration is unstable to an ideal MHD instability with Alfv\'enic growth rates, and we will vary $B_{z0}$ to model cases at different prevailing plasma beta conditions. When we set the density $\rho$ to unity initially, and fix $p_0=1/\gamma$ for a ratio of specific heats $\gamma=5/3$, the implied normalization uses the sound speed external to the double current system as unit of speed, the radius of the double current channel as unit of length, and the density fixes our unit of mass. 

We perturb the equilibrium with an incompressible velocity field given by
\begin{eqnarray}
v_x & = & +\frac{\partial \phi_0}{\partial y} \,\,\, \left[ \times \sin(k_z z) \right] \,, \nonumber \\
v_y & = & -\frac{\partial \phi_0}{\partial x} \,\,\, \left[ \times \sin(k_z z) \right] \,, \nonumber \\
v_z & = & 0 \,, 
\label{q-vval}
\end{eqnarray}
where the streamfunction $\phi_0(x,y)=\epsilon \exp(-x^2-y^2)$ has amplitude $\epsilon=0.0001$. The dependence on the third $z$-coordinate only applies to the 3D cases, where the vertical box size $L_z=2\pi/k_z$. When performing 3D runs, we fixed $L_z=6$ and also take $z\in [-3,3]$. 

In all runs, we integrate the standard set of resistive, compressible 3D MHD equations, where we use finite, uniform and constant resistivity parameter $\eta=0.0001$. It was pointed out by~\citet{richard90} that due to the ideal nature of the instability, the resistivity has little effect on the linear phase, but clearly influences the nonlinear stage and the possibilities for reconnection. We use varying grid resolutions to study how numerical versus actual resistivity values interplay in our evolutions. A recent study by~\citet{popmsdg} has highlighted how various modern discretizations handle especially the chaotic reconnection regime at high magnetic Reynolds numbers, where secondary tearing events dominate, and part of our results below are surely influenced by discretization aspects mentioned in that study.

For the boundary conditions, when 3D runs are done, the $z$-direction is periodic. The lateral $(x,y)$ boundaries are handled by continuous extrapolation of the primitive variables $\rho$, $\mathbf{v}$ and $p$ from the closest inner mesh cell value into all ghost cells (i.e. zero gradient is adopted).
The magnetic field in the ghost cells first fixes the analytic profiles from the original equilibrium variation (e.g. $B_x=2xy/r^4$), and subsequently exploits a second order, central finite difference evaluation of the divergence of the magnetic field to correct the component normal to the boundary at hand. Overall, this prescription works well for full grid-adaptive runs, in combination with a diffusive approach on the monopole error control. There, the same discretization is used to quantify $\nabla\cdot \mathbf{B}$, which is then added as a diffusion part to the induction equation in the form $\nabla\left((\Delta x)^2 \nabla\cdot \mathbf{B}\right)$, as one among several options available in the open-source MPI-AMRVAC software~\citep{porth14,amrvac12}. For the spatiotemporal advance, we here use a three-step Runge-Kutta type scheme, with a Harten-Lax-van Leer (HLL) flux prescription and a third order limiter~\citep{cada,kepjcam14}.

\begin{table}
\begin{center}
\begin{tabular}{|c|ccc|cc|c|}
\hline
\vphantom{\LARGE B}
run  & $B_{z0}$ & $\bar{\beta}, \beta_\infty$ & effective resolution & $\bar{j}$ & $\bar{T}$ & $\gamma_{\mathrm{tilt}}$ \\
\hline
\hline
\vphantom{\LARGE B}
A2d & $0.0$ & $12.7, 1.20$ & $2400\times 2400$ & $\pm 4.35$ & $1.6$ & 1.4978 \\ 
\vphantom{\LARGE B}
b2d & $0.1$ & $5.8, 1.19$ & $300\times 300$ & $\pm 4.35 $ & $1.6$ & \\
\vphantom{\LARGE B}
B2d & $0.1$ & $5.8, 1.19$ & $2400\times 2400$ & $\pm 4.35 $ & $1.6$ & 1.4977 \\
\vphantom{\LARGE B}
BB2d & $0.1$ & $5.8, 1.19$ & $4800\times 4800$ & $\pm 4.35 $ & $1.6$ &  \\
\vphantom{\LARGE B}
C2d & $0.5$ & $2.6, 0.96$ & $2400\times 2400$ & $\pm 4.35 $ & $1.6$ & 1.4903 \\
\vphantom{\LARGE B}
D2d & $1.0$ & $1.4, 0.60$  & $2400\times 2400$ & $\pm 4.35 $ & $1.6$  & 1.4809 \\
\vphantom{\LARGE B}
E2d & $5.0$ & $0.12, 0.046$  & $2400\times 2400$ & $\pm 4.35 $ & $1.6$  & 1.3556 \\
\hline
\vphantom{\LARGE B}
A3d & $0.0$ & $12.7, 1.20$ & $300^3$ & $\pm 4.35$ & $1.6$ & \\
\vphantom{\LARGE B}
B3d & $0.1$ & $5.8, 1.19$ & $300^3$ & $\pm 4.35 $ & $1.6$ & \\
\vphantom{\LARGE B}
BB3d & $0.1$ & $5.8, 1.19$ & $600^3$ & $\pm 4.35 $ & $1.6$ & \\
\vphantom{\LARGE B}
C3d & $0.5$ & $ 2.6, 0.96$ & $300^3$ & $\pm 4.35 $ & $1.6$ & \\
\vphantom{\LARGE B}
D3d & $1.0$ & $1.4, 0.60$  & $300^3$ & $\pm 4.35 $ & $1.6$  & \\
\vphantom{\LARGE B}
DD3d & $1.0$ & $1.4, 0.60$  & $600^3$ & $\pm 4.35 $ & $1.6$  & \\
\vphantom{\LARGE B}
EE3d & $5.0$ & $0.12, 0.046$  & $600^3$ & $\pm 4.35 $ & $1.6$ &  \\
\hline
\end{tabular}
\caption{The simulated cases and several characteristic parameters. The leftmost column serves to label the various runs, the right column quantifies tilt mode growthrates (see text for details).}
\label{t-cases}
\end{center}
\end{table}
The 2D runs all have minimal resolution $300^2$, but when using 4-5 AMR grid levels can achieve $2400^2-4800^2$ effective resolution. In the 3D setups, we explored grid sizes from $300^3$ to $600^3$, where we used either domain decomposition on fixed grid sizes, or 3 AMR levels from $150^3$ base grids. In Table~\ref{t-cases}, the most important parameters quantifying the various cases are listed. We give representative mean initial values for the prevailing plasma beta $\bar{\beta}$, current density $\bar{j}$ and temperature $\bar{T}$ (these two quantities are identical for all runs), for differing choices of $B_{z0}$, where the mean value for a scalar $f$ is computed over the current channel cross-section as follows
\begin{equation}
\bar{f}^{\pm}\equiv \frac{\iint_{j_z(t=0)\stackrel{>}{<}0} f \,dx\,dy }{\iint_{j_z(t=0)\stackrel{>}{<}0} \,dx\,dy} \,.
\label{q-mean}
\end{equation}
The $\pm$ superscript differentiates between positive versus negative current density $z$-component, and at $t=0$ we find the denominator area value to be $\pi/2$. For any later time, we can no longer purely depend on the sign of $j_z(t)$, since reconnection and other dynamic events cause mixed distributions throughout. In order to identify the location of the initially purely positive or negative current channels at all later times, we additionally advect a tracer that identifies their displaced location. This tracer will be used further on to quantify the current channel displacements, or to quantify energetics specific to each of both evolving current channels. Table~\ref{t-cases} also lists the asymptotic plasma beta, computed from $\beta_\infty=2 p_0/(1+B_{z0}^2)$.

\section{Results in 2.5D configurations}\label{s-25d}

\begin{figure}
\begin{center}
\FIG{
\resizebox{0.49\textwidth}{8cm}
%{\includegraphics{figures/tiltioFigAtr5}}
{\includegraphics{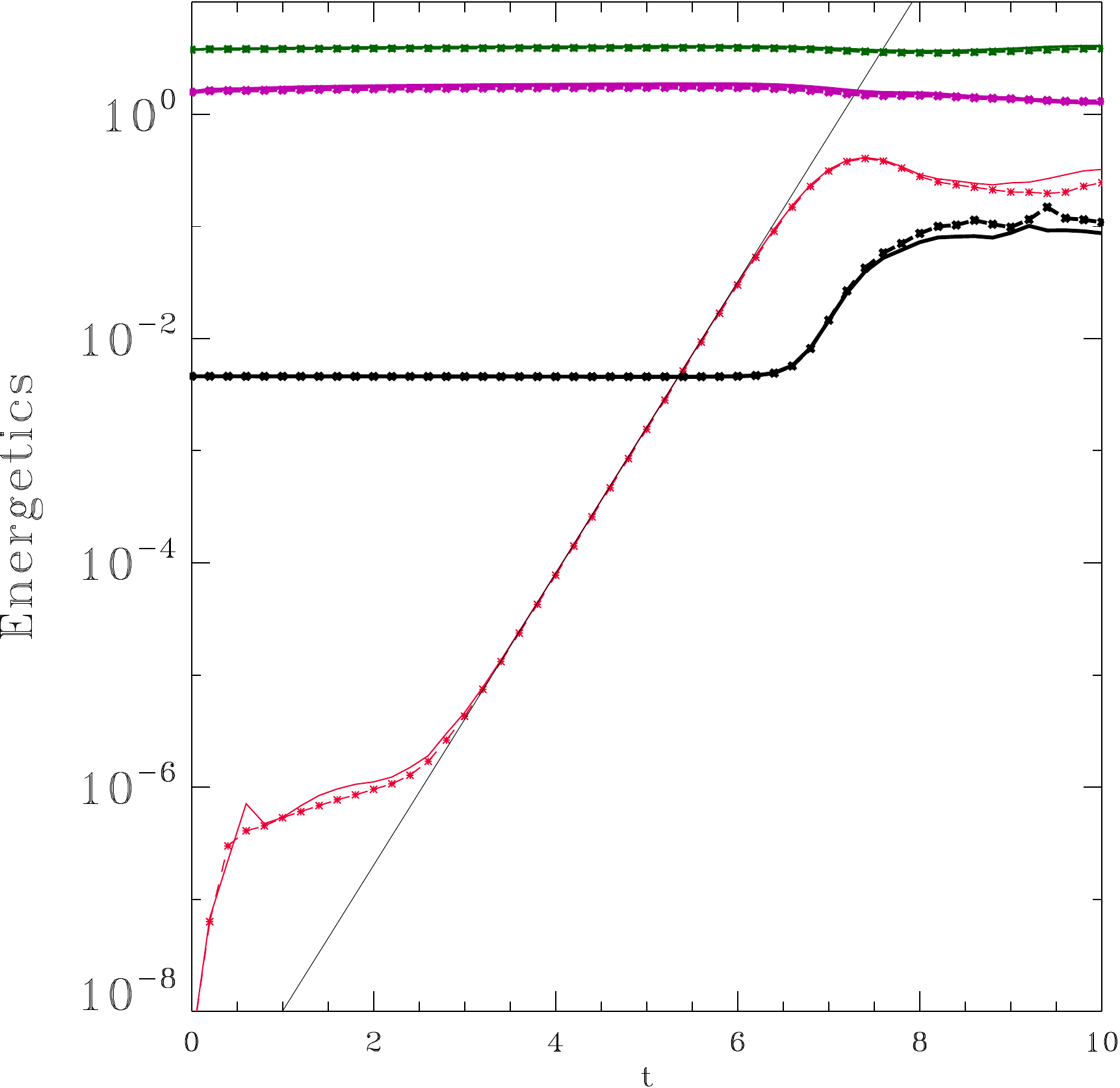}}
\resizebox{0.49\textwidth}{8cm}
%{\includegraphics{figures/runtiltresmhd23}}
{\includegraphics{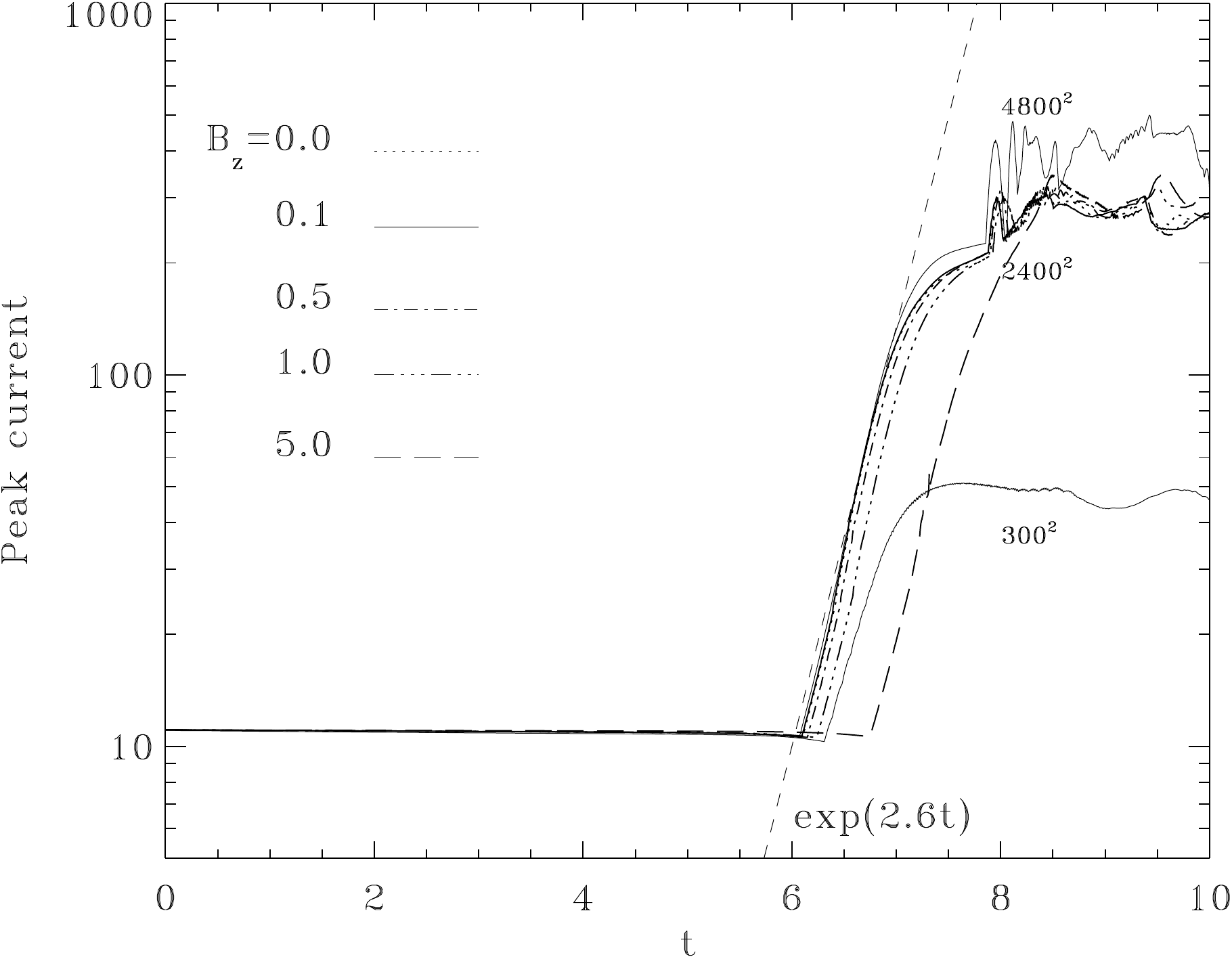}}
}
\end{center}
\caption{Left panel: For a case with $B_z=0.1$, the evolution of kinetic (red), magnetic (purple) and internal energy (green) density, at all times integrated over a single current channel as identified by an advected tracer. We show each curve at two resolutions (higher resolution with connected symbols). The black curve quantifies the Ohmic heating effect. The thin black straight line is a fit to the linear growth in kinetic energy. Right panel: the evolution of the peak current, for all 2.5D runs with linestyles distinguishing different axial field cases. Different resolutions are also shown for $B_z=0.1$. Exponential growth at $\exp(2.6t)$ is indicated to guide the eye.}
\label{fig1}
\end{figure}

As already mentioned in the introduction, the ideal MHD equilibrium where two adjacent current channels are in a force-balanced state is subject to an ideal MHD instability that wants to mutually repel and seperate the antiparallel current channels. The linear instability in this essentially 2D configuration contains both an antiparallel displacement of each current channel along the $y$-direction ($x=0$ separates the channels within the unit circle initially), as well as a rotational, twisting motion. An analytic stability analysis exploiting the energy principle (assuming incompressible conditions) showed that pure rotation is only marginally stable, while the combination of displacement and rotation can drive instability~\citep{richard90}. \cite{richard90} quantified growth rates by monitoring the total bulk flow kinetic energy evolution, where a clear linear growth phase could always be identified. We use a similar means to quantify this growth rate, but instead of using the kinetic energy over the total simulation box, we use the tracers mentioned earlier to quantify energy measures for the individually displacing current channels. In particular, the left panel of Fig.~\ref{fig1} quantifies $\tilde{f}(t)\equiv \iint_{\mathrm{tr}>0} f \,dx\,dy$ for kinetic energy density $f=0.5 \rho v^2$, as well as magnetic energy density $0.5 B^2$, internal energy density $p/(\gamma-1)$ and Ohmic heating term $\eta J^2$ (noting that the latter is really an energy density rate change, with different units: code units are used throughout all plots). This panel actually contains two curves for each quantity, as we plot both case B2d and BB2d (this higher resolution run is shown with connected symbols), to demonstrate nicely converged results as judged from these global energetic indicators. The perfect linear growth phase in kinetic energy spans the entire time interval $t\in [3,7]$. To quantify this growth, we used a linear fitting routine for data with $t\in[4,6]$, yielding the indicated linear fit, and half its slope sets the growth rate $\gamma_{\mathrm{tilt}}^B=1.4977$. We verified that this growth rate estimate is the same when using the other current channel, or when using the entire domain kinetic energy. This case B2d has internal energy dominating magnetic energy (due to the $\beta>1$ conditions), both clearly dominating kinetic energy at all times. The evolution of the Ohmic heat term also shows a transition at a time around $t\approx 6.5$ that lags the linear tilt mode growth phase. This is caused by the fact that in the nonlinear, saturation stage of the tilt deformation, in essence near-singular current sheets develop on the leading front of both rotating channels, where antiparallel field line regions meet. This phase is numerically challenging, and has been used as a stringent test to test adaptive finite element methods for incompressible MHD~\citep{lankalapalli07,strauss98}, where it became clear that a phase where the {\em logarithm} of the peak current density grows linearly gets established. To follow this linear growth for $\log(\max{j_z})$ over more than one order of magnitude required (combinations of) $h$- and $p$-refinement strategies. Our block-AMR strategy, together with our third-order spatio-temporal treatment, allows to resolve this phase accurately, untill secondary tearing type disruptions cause a more chaotic regime with chaotic island formations on the disrupting current sheets. The right panel of Fig.~\ref{fig1} shows the evolution of the peak current in a log-linear scale for all 2.5D cases. For the case with $B_z=0.1$ we show all three resolutions (b2d, B2d, BB2d), and note that this more stringent local measure of `convergence' as yet shows differences in the island-dominated phase beyond $t\approx 8$ for the highest resolutions. This observation is consistent with findings on the double periodic GEM-type reconnection setup explored with different discretization strategies in~\citet{popmsdg}, focusing on the chaotic secondary island formation regime. In Fig.~\ref{fig1}, the $300^2$ run is seen to reach much lower peak current values, while the $2400^2$ and $4800^2$ run agree throughout most of the linear growth phase for $\log(\max{j_z})$. Comparing the runs at different $\beta$ and identical resolution $2400^2$, one notices that a systematic delay is seen in the onset of this singular current development when lowering $\beta$. At the same time, all cases roughly agree on the slope for this singularity onset, with an indicative $\log(\max{j_z})\sim 2.6 t$ as plotted by the dashed line. While this nonlinear near-singularity development is hence similar for all cases, the systematic delay in its onset is consistent with the fact that when we quantify tilt mode growth rates as mentioned previously for case B2d with $B_z=0.1$, we find a decreasing growth rate as the plasma $\beta$ decreases from cases A2d to E2d, namely
$\gamma_{\mathrm{tilt}}^A=1.4978$, $\gamma_{\mathrm{tilt}}^B=1.4977$, $\gamma_{\mathrm{tilt}}^C=1.4903$, $\gamma_{\mathrm{tilt}}^D=1.4809$, and $\gamma_{\mathrm{tilt}}^E=1.3556$. We note that this trend is opposite to the one reported by~\citet{richard90}, where pure 2D force-free (and hence constant plasma pressure) conditions in plasma beta ranges between $0.08$ to $2$ were investigated. The difference can be due to this difference in initial setup, although we note that also the reduced resolution (a factor 10 lower than ours) and their use of a more diffusive second order scheme with added purely numerical diffusion terms may interfere. Note that the growthrates mentioned above use our dimensionalization exploiting sound crossing times, we can translate to Alfv\'enic crossing time units by multiplying with the factor $c_s/v_A=1/\sqrt{1+B_{z0}^2}$ (using values for sound speed $c_s$ and Alfv\'en speed $v_A$ at far distances), and this gives
$\gamma_{\mathrm{tilt,a}}^A=1.4978$, $\gamma_{\mathrm{tilt,a}}^B=1.4903$, $\gamma_{\mathrm{tilt,a}}^C=1.3329$, $\gamma_{\mathrm{tilt,a}}^D=1.0471$, and $\gamma_{\mathrm{tilt,a}}^E=0.2658$.

\begin{figure}
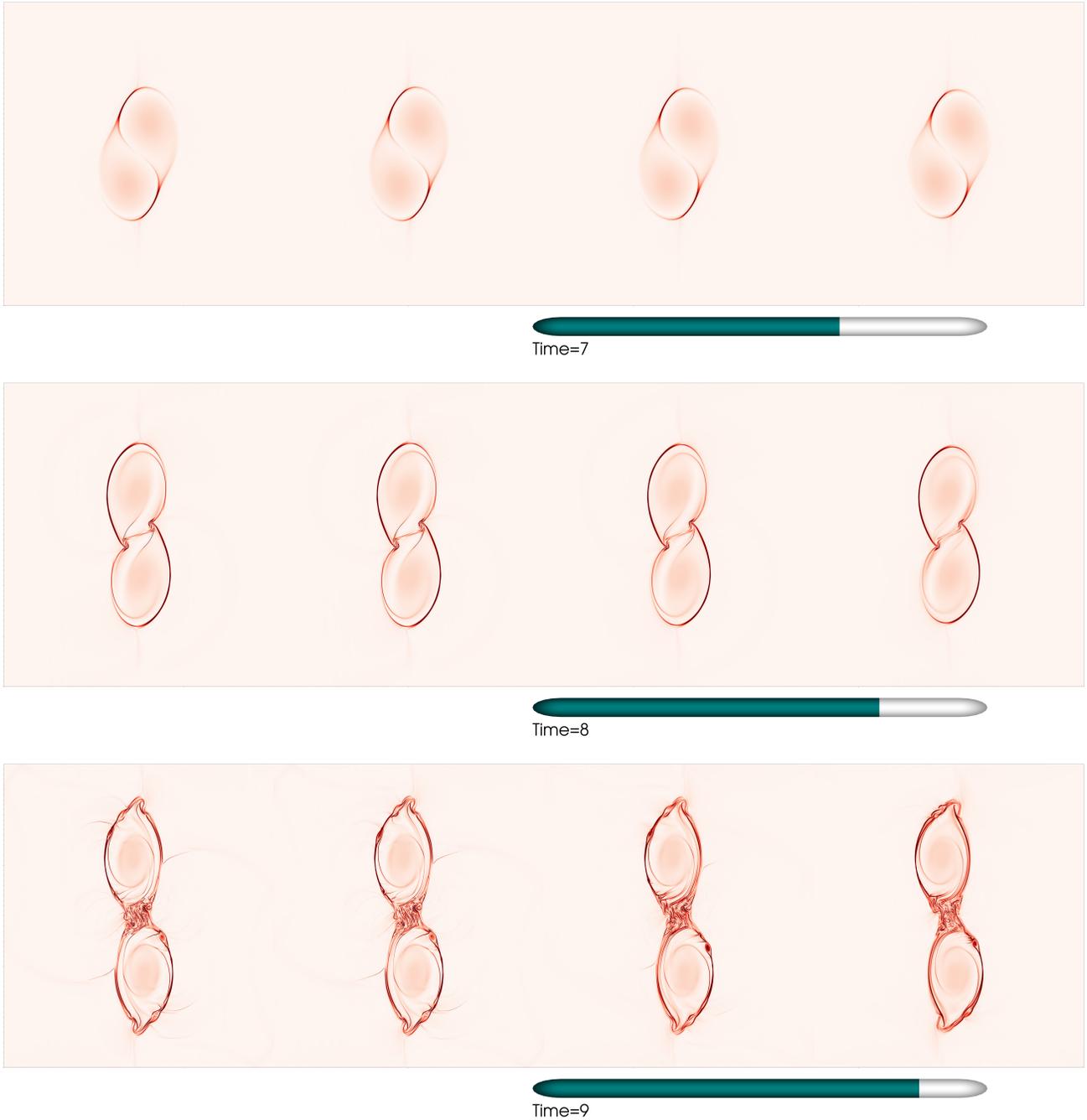

\begin{center}
\FIG{
\resizebox{\textwidth}{!}
%{\includegraphics{figures/fig_tilt23t7}}
{\includegraphics{fig2a}}
\resizebox{\textwidth}{!}
%{\includegraphics{figures/fig_tilt23t8}}
{\includegraphics{fig2b}}
\resizebox{\textwidth}{!}
%{\includegraphics{figures/fig_tilt23t9}}
{\includegraphics{fig2c}}
}
\end{center}
\caption{The evolution of the total current magnitude, for cases A2d (left) through D2d (right) with $B_z$ increasing from left to right. The three rows correspond to times $t=7$, 8, and 9, respectively. A linear color scale is saturated to show values between $[0-50]$, showing all structure.}
\label{fig2}
\end{figure}

\begin{figure}
\begin{center}
\FIG{
\resizebox{0.5\textwidth}{!}
%{\includegraphics{figures/tiltioFigAtr6}}
{\includegraphics{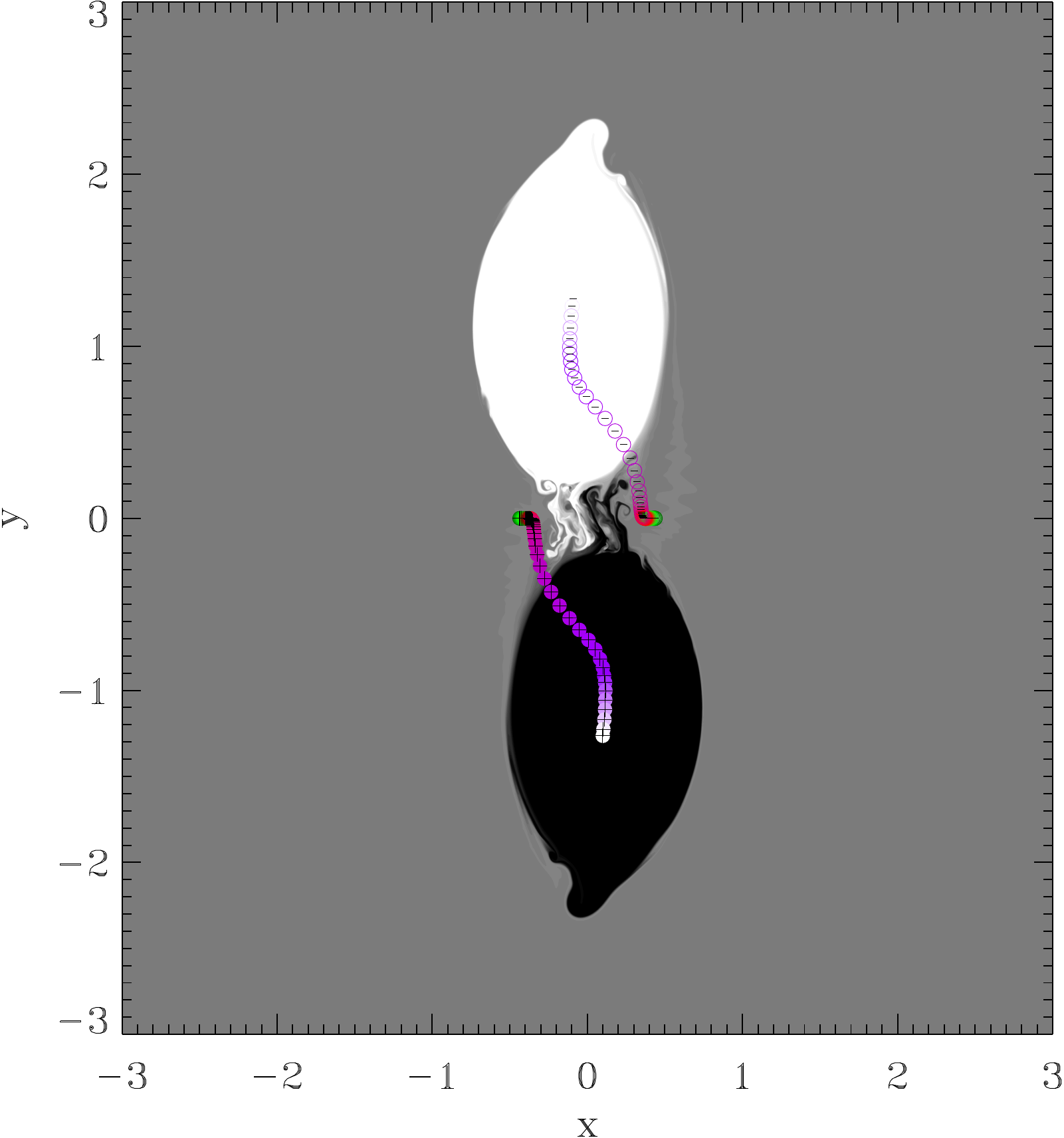}}
}
\end{center}
\caption{For case BB2d, with $B_z=0.1$, this figure shows the tracer distribution used to follow the current channel movements at all times, here at time $t=9$. The colored symbols denote the central location of the positive ($+$ symbols and black tracer region) and negative ($-$ symbols and white tracer regions) current channels, over all times $t\in[0,10]$.}
\label{fig3}
\end{figure}

Despite the weak plasma beta dependence on the linear tilt mode growthrate, all cases ultimately show the singular current growth and they saturate in a very similar manner, with peak currents reached of order ${\cal{O}}(300)$, a factor 30 above the initial peak current value. The overall evolution for all five cases A2d to E2d are similar, and this is made visible for cases A2d through D2d in Fig.~\ref{fig2}. This figure shows these four cases next to each other at three consecutive times $t=7$, 8 and 9, plotting the total current magnitude. To show all structure and not purely the near singular current peak values, we used the same linear color scale for all frames, artificially restricted to values between $0-50$. It clearly shows the near-identical evolution for all 2.5D cases (case E2d is not shown, but similar), and that especially in the later stages beyond $t=8$ one witnesses the disruption of the singularly enhanced current sheets by tearing type chaotic reconnection. The presence of a magnetic field component perpendicular to the plane shown for cases B2d through E2d does lead to detailed differences in where these islands first appear, and how they induce wave fronts exterior to the deforming islands. In the region between the seperating current channels, a fairly turbulent region develops where the plasma from the original current channels effectively mixes. This is seen best when plotting the tracer quantity that is used to identify the displaced current channel matter, and Fig.~\ref{fig3} shows this tracer at $t=9$ for the highest resolution run BB2d. Indications of both interchange and shear-flow type deformations can be detected. The figure also plots the path taken by the originally purely negative (indicated with $-$ symbols and corresponding to the white tracer region) and originally purely positive ($+$-symbols and the black tracer region) current channels, by at all times locating the $x$ and $y$-moment values for both tracer regions, i.e. determining $x_{\mathrm{tr}}=\tilde{x}(t)/\tilde{1}(t)$, $y_{\mathrm{tr}}=\tilde{y}(t)/\tilde{1}(t)$. This $(x_{\mathrm{tr}},y_{\mathrm{tr}})$ path visually shows the displacement experienced by each current channel due to the tilt instability, also well into the nonlinear phase.

\section{Results for 3D configurations}\label{s-3d}

\begin{figure}
\begin{center}
\FIG{
\resizebox{0.5\textwidth}{!}
%{\includegraphics{figures/runtiltresmhd33}}
{\includegraphics{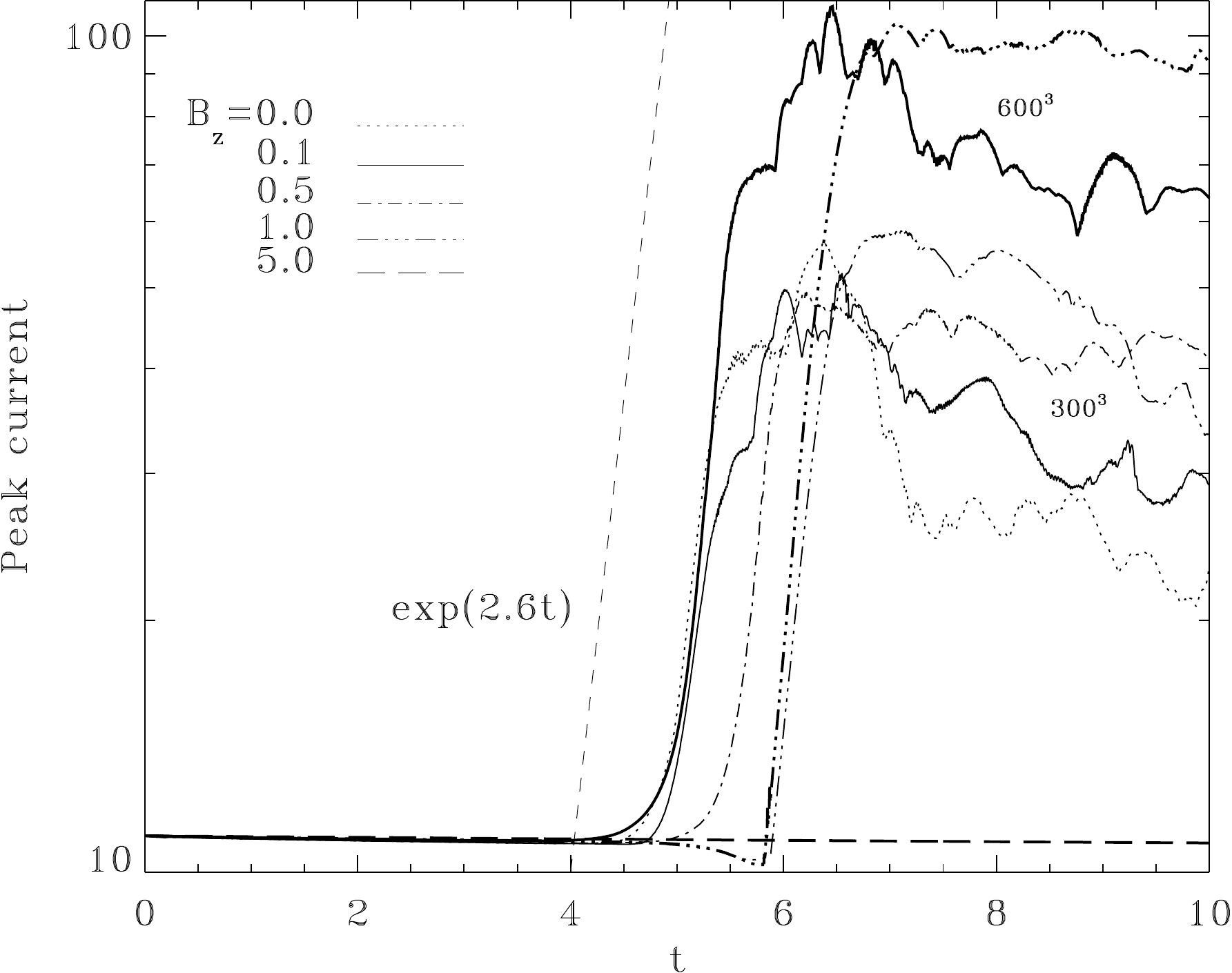}}
}
\end{center}
\caption{The peak current evolution for all 3D cases, with the exponential curve as a dashed line to guide the eye.}
\label{fig4}
\end{figure}

The 2.5D setup did not show significant differences over the entire range of plasma beta, and the role of a vertical $B_z$ component was minimal since the translational invariance prevented potentially stabilizing field line bending. In a true 3D setup, two additional effects come into play, the first being that field lines may bend w.r.t. the $z$-direction, and the fact that each individual current channel may well be unstable to ideal kink deformation. When we quantify the ratio
$K_{\mathrm{cr}}\equiv |\tilde{j_z}|/\tilde{B_z}$ from the initial condition for cases B2d through E2d, we find respectively 43.5, 8.7, 4.35, 0.87. Taken together with case A2d, where this ratio is formally infinite due to the vanishing $B_z$, these values demonstrate a clear liability to kink deformations for all except case E2d. This is judged from the Kruskal-Shafranov limit where $K_{\mathrm{cr}}< 2 a/R_0 = 4 \pi a/L\approx 1$ denotes stability, where we use a plasma column radius $a\approx 0.5$ and length $L=6$. Although the configuration is not the idealized plasma-vacuum setup from the Kruskal-Shafranov analysis, we expect all but case EE3d to have an insufficient magnetic field $B_z$ component to stabilize kink deformations.

Figure~\ref{fig4} shows the obtained evolution of the peak current for all 3D runs (similar to Fig.~\ref{fig1}, right panel for the 2.5D cases). The dashed line serves to guide the eye, and quantifies the same exponential growth $\exp(2.6 t)$ we indicated in Fig.~\ref{fig1} for the peak current. Two trends are evident: as the $B_z$ component increases from model runs A3d ($B_z=0$) through D3d ($B_z=1$), the onset of this near-singular peak current evolution happens later, to find full suppression of any instability development for the case EE3d ($B_z=5$) which maintains its peak current near the initial value. Indeed, this case EE3d has a sufficiently strong $B_z$ component to ensure both suppression of potential kink disruptions internal to both channels as argued above, but also to prevent the full development of tilt instability between the repelling current channels (at least during the simulated timeframe). Tilt evolution is, due to the $\sin(2\pi z/L_z)$ dependence on the imposed velocity perturbation, in principle expected at both locations $z=\pm L_z/4$ where the 3D setup behaves near-identical to the 2.5D configuration. However, the now included stabilizing influence of field line bending prevents the tilt development in the present setup. We verified with a seperate simulation identical to EE3d, but where the initial velocity perturbation has no $z$-dependence at all (replacing the factor $\sin(2\pi z/L_z)$ by unity in Eqns.~\ref{q-vval}) that in this case, the 3D behaves identical to the 2.5D case, and does indeed demonstrate tilt development. This confirms the role of magnetic tension as a stabilizing actor. We expect that configurations with such strong axial field components may only become tilt unstable for larger box sizes $L_z$, at correspondingly longer wavelengths.

Another aspect demonstrated in Fig.~\ref{fig4} is that for the two runs where both a $300^3$ and a $600^3$ resolution is shown, the peak current evolution is similar, except for nonlinear saturation levels attained. This is not surprising given the similar findings in the even higher resolution 2.5D runs discussed, where we again stress that more global convergence measures and visual data inspection confirm that sufficient detail is already captured at $300^3$.

\begin{figure}
\begin{center}
\FIG{
\resizebox{\textwidth}{!}
%{\includegraphics{figures/fig_tilt33Bz01t6}}
{\includegraphics{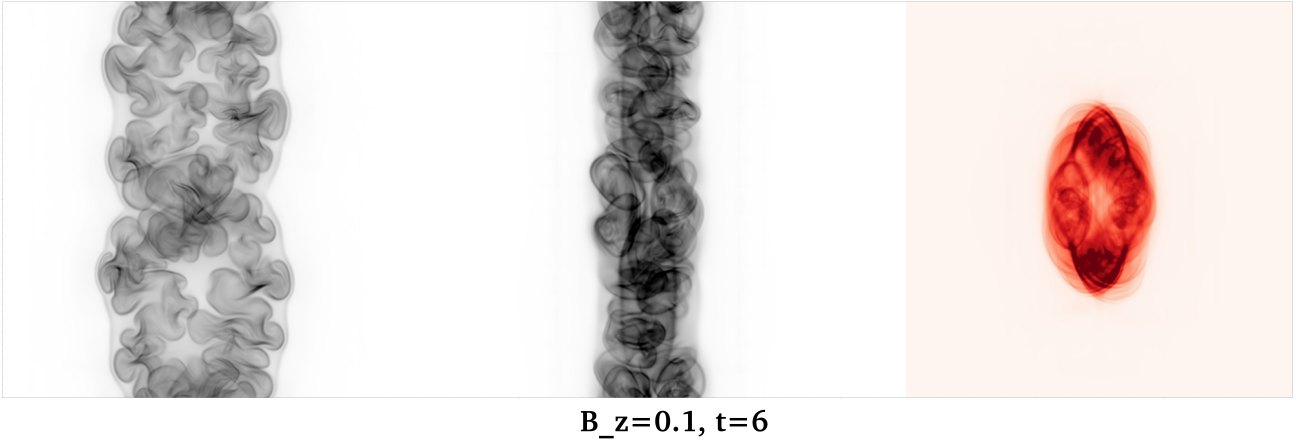}}
}
\end{center}
\caption{The total current, integrated along the line of sight, for the highest resolution $B_z=0.1$ case at $t=6$. At left, we integrate along $x$, with color scale saturated to 40. In the middle, we integrate along $y$, and saturate at 40. At right, we integrate along $z$, and use saturation at 80.}
\label{fig5}
\end{figure}

\begin{figure}
\begin{center}
\FIG{
\resizebox{\textwidth}{!}
%{\includegraphics{figures/fig_tilt33Bz05t6}}
{\includegraphics{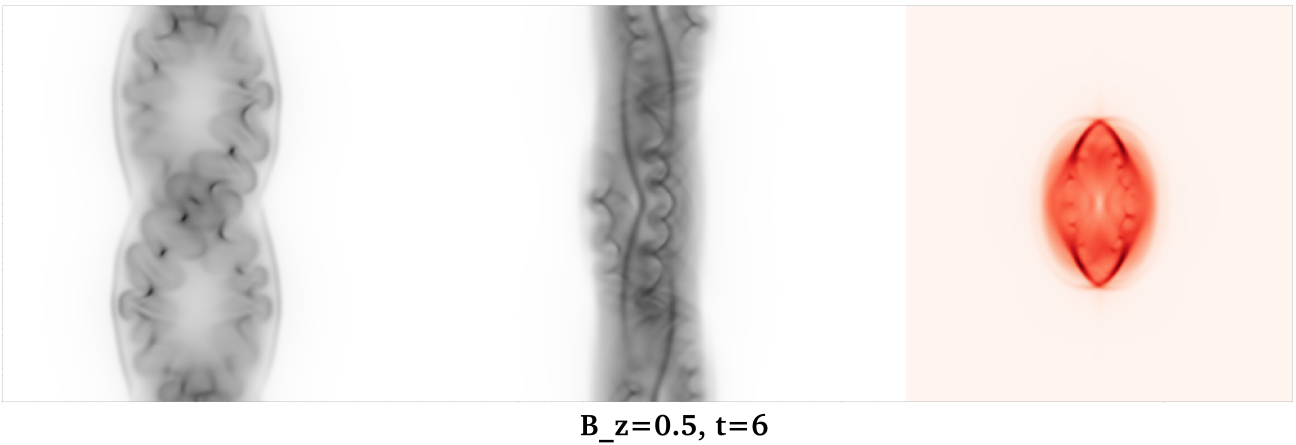}}
}
\end{center}
\caption{As in figure~\ref{fig5}, for case $B_z=0.5$. The color scales are now saturated at 30, 30, and 70 from left to right.}
\label{fig6}
\end{figure}

Turning attention to the 3D runs which do develop tilt-driven peak-current enhancement, all cases with $B_z$ increasing from 0 to 1 show additionally rich fine-structure developing due to kink deformations within both current channels. Figures~\ref{fig5}-\ref{fig6} show representative views for two cases (from case BB3d and C3d, respectively) where we show line-integrated views on the total current magnitude (using the collapse view option as described in~\citet{porth14}). Shown is a contour representation of $\int_{-3}^{+3} j \,dx_i$, from two lateral views (i.e. $x_i=x$ and $x_i=y$) and one top view ($x_i=z$), from left to right. All frames are taken at time $t=6$ when the peak current saturation is nearly complete, and the contour level range is adapted from frame to frame as indicated in the figure caption, to enhance all details. In both figures, the two lateral views show helical substructure which corresponds to the kinking of both current channels. Higher axial field components (contrasting Fig.~\ref{fig5} to Fig.~\ref{fig6}) suppresses some, but not all of the secondary mushroom-like features seen especially at far left ($x$-integrated view). Animated views reveal clearly the tilt displacement (connected to the two-bulged pattern seen in that view) where both current channels repel one-another and get displaced primarily along the $y$-direction. The middle, $y$-integrated views by now strongly mix contributions from both channels (in accord with the in-plane orientation as in our Figure~\ref{fig3}). The top views show the outlines of the most strongly peaked current concentrations, in line with the expected patterns seen in the 2.5D cases from Figure~\ref{fig2}. 

\begin{figure}
\begin{center}
\FIG{
\resizebox{0.49\textwidth}{!}
%{\includegraphics{figures/fig_tilt33Bz01t6j10rho3dview}}
{\includegraphics{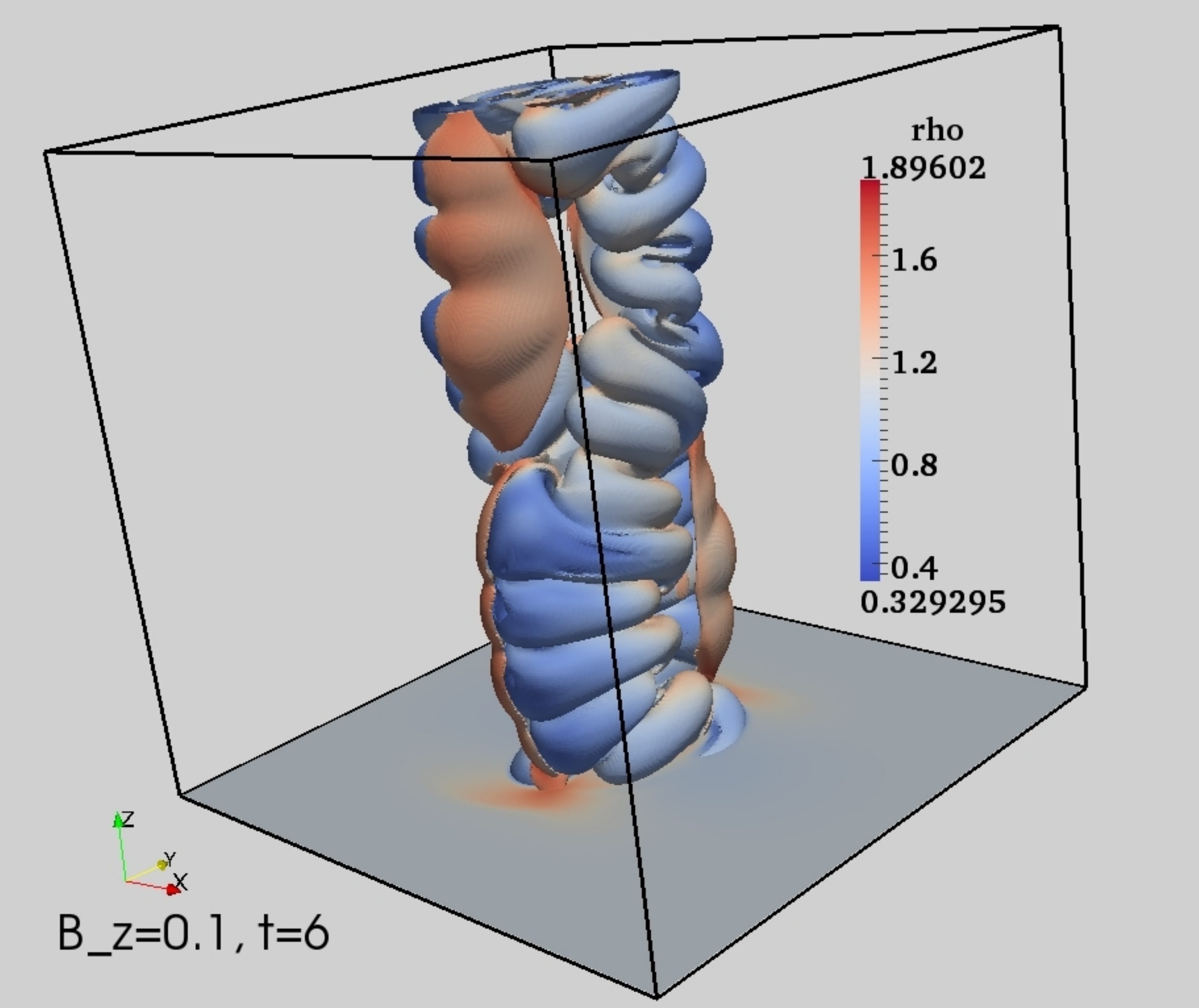}}
\resizebox{0.49\textwidth}{!}
%{\includegraphics{figures/fig_tilt33Bz05t6j10rho3dview}}
{\includegraphics{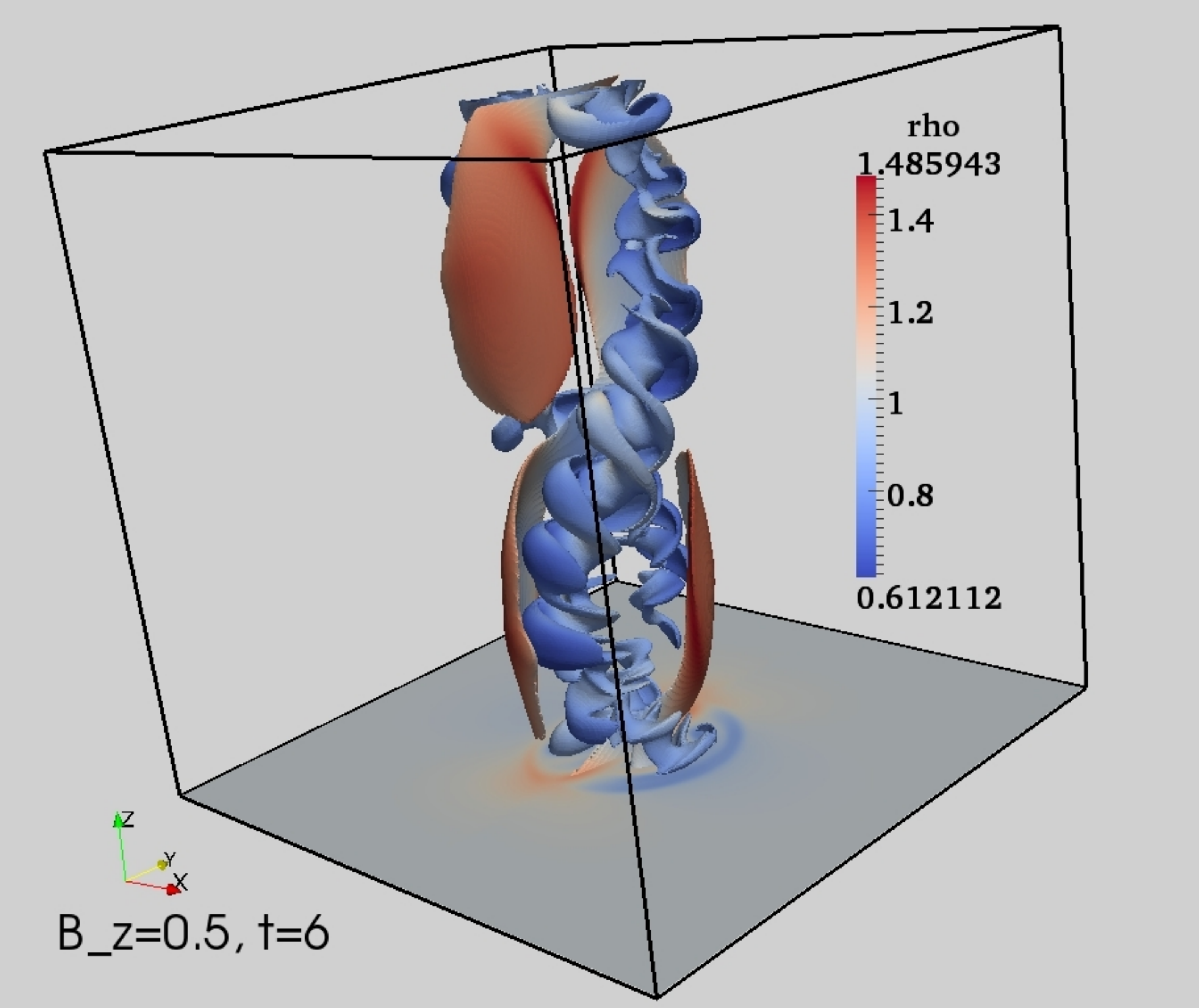}}
}
\end{center}
\caption{For the same cases and times as shown in Fig.~\ref{fig5}-\ref{fig6}, this 3D view shows the total current isosurface where $j=10$, colored by density. Also the bottom plane shows the density distribution, all at times $t=6$.}
\label{fig7}
\end{figure}

\begin{figure}
\begin{center}
\FIG{
\resizebox{0.5\textwidth}{!}
%{\includegraphics{figures/runCt6fieldEBnew}}
{\includegraphics{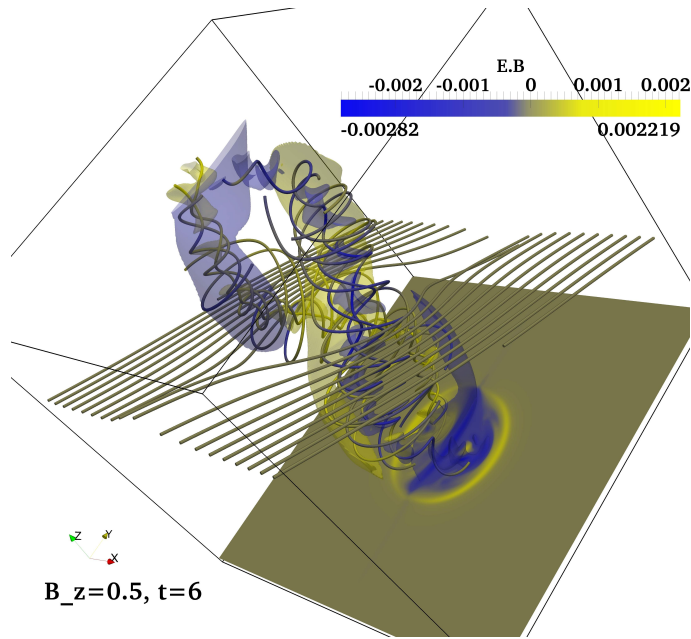}}
}
\end{center}
\caption{For the case with $B_z=0.5$ as shown in Fig.~\ref{fig6}-\ref{fig7}, we show the magnetic field structure, and the $\mathbf{E}\cdot\mathbf{B}$ variation, using translucent isosurfaces at values $\pm 0.001$, accordingly colored field lines, and its bottom plane contour view.}
\label{fig8}
\end{figure}

\begin{figure}
\begin{center}
\FIG{
\resizebox{0.5\textwidth}{!}
%{\includegraphics{figures/runCt9j10ebB}}
{\includegraphics{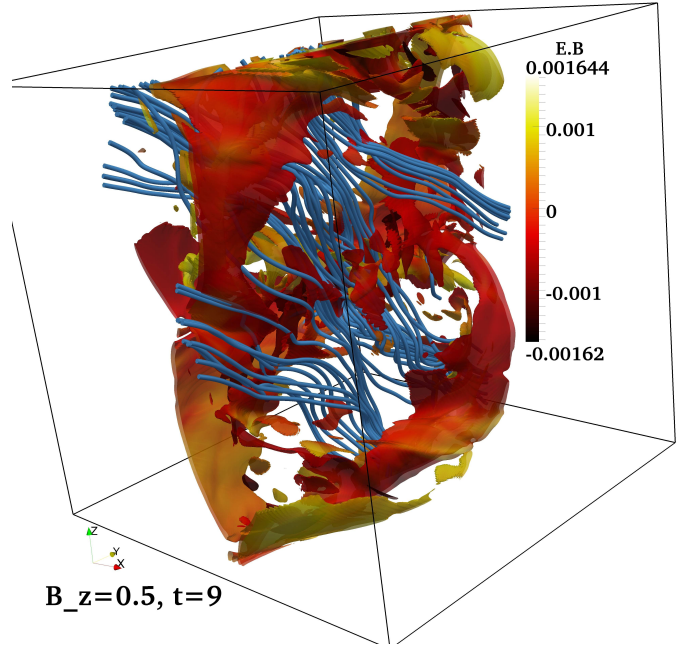}}
}
\end{center}
\caption{The same case shown in Fig.~\ref{fig8}, now in the far nonlinear disruption regime at $t=9$. Shown is the $j=10$ isosurface previously displayed in Fig.~\ref{fig7}, but here colored by the local $\mathbf{E}\cdot\mathbf{B}$ value, and selected field lines (in blue), all passing through the central line $x=0=y$.}
\label{fig9}
\end{figure}

The line integrated views are complemented with true 3D renderings of the current concentrations for the same cases and time $t=6$ in Figure~\ref{fig7}. This figure shows the 3D shape of the total current isosurface where $j=10$, which is colored by the local density value as indicated. Also the bottom $z=-3$ plane shows the density variation. Both cases demonstrate the clearly kink-related helical deformation of the current concentration in each channel. Sheetlike structures are found in four locations, whithin which we find the peak current amplifications due to tilt deformation. These sheetlike structures are enhanced in density (reddish color), as material gets compressed due to the tilt deformations. A further 3D view on the $B_z=0.5$ case C3d at the same time $t=6$ as displayed in Figures~\ref{fig6}-\ref{fig7} shows an impression of the field line structure in Figure~\ref{fig8}. In this view, we also indicated the instantaneous distribution of the scalar product of electric and magnetic field, i.e. $\mathbf{E}\cdot\mathbf{B}=\eta\mathbf{j}\cdot\mathbf{B}$. In the figure, the variation of this quantity is shown on a horizontal cut, but also in isosurface views: the translucent yellow and blue isosurfaces correspond to values $0.001$ and $-0.001$, respectively. Also the field lines are colored by their local $\mathbf{E}\cdot\mathbf{B}$ value. As is well-known from basic MHD theory, locations where 3D reconnection can occur and ideal MHD assumptions break down are diffusion regions where $\int_{fl}\mathbf{E}_{\parallel} \cdot \mathbf{dl}\ne 0$, i.e. where electric fields parallel to the magnetic field lines develop~\citep{schindler88}. The shown $\mathbf{E}\cdot\mathbf{B}$ variation confirms that likely reconnection sites develop in the strong current layers, but also in between the repelling current channels, and internal to the kink-deforming current channels. Where anti-parallel field lines meet, diffusion regions develop and topological rearrangements occur. In fact, the saturation phase that follows this time $t=6$ snapshot shows very small-scale, but highly structured disruptions happening throughout the tilt-induced separating current channels, mitigated by the internal kink evolutions. An impression of this is shown in Figure~\ref{fig9}, where for the same case C3d (with $B_z=0.5$), we show the same total current $j=10$ isosurface previously displayed for $t=6$ in the right panel of Figure~\ref{fig7}. This time, it is colored by the local $\mathbf{E}\cdot\mathbf{B}$ variation. Representative field lines, all passing through the central line $x=0=y$ are shown as well. It is then obvious how fairly global topological rearrangements have established field connections between back and front sided (i.e. $x>0$ and $x<0$) regions throughout the disrupting current channels. A significant amount of localized, strong current sheets are evident. Most notable are still the four large current sheets where the peak currents are located. Their deformation still follows the trend where top and bottom ($z>0$ versus $z<0$) parts of each current channel get displaced in opposite directions, along the $y$-coordinate lines. Similar observations hold for all other cases where combined tilt and kink instability developments set in (i.e. all but case EE3d).

\section{Discussion and solar coronal application}\label{s-disc}

The main findings of our study can be summarized as follows. First, purely planar tilt evolutions, as well as translationally invariant extensions of the basic setup where two neighbouring islands carry opposite currents, were studied in compressible MHD. As in the previous incompressible planar~\citep{strauss98,lankalapalli07}, or compressible force-free~\citep{richard90} scenarios, the ideal tilt manifests itself over a wide range of plasma beta ($\bar{\beta}\in[0.12,12.7]$ have been explored, see table~\ref{t-cases}) and has Alfv\'enic growth rates, which decrease for decreasing $\beta$. Using extreme resolution, adaptive simulations, we showed that the near singular current sheets that develop ahead of the tilt-displaced islands additionally become liable to tearing-type disruptions. The linear growth for the logarithm of the peak current concentration ultimately saturates due to these disruptions, although we could demonstrate its growth for close to two orders of magnitude.
Secondly, full 3D scenarios are enriched by the possibility to kink-deform both current channels, during the tilt development. Contrary to the translationally invariant cases, the axial magnetic field component can play a stabilizing role, for both the kink and the tilt instability. Tilt-kink unstable double current channel configurations will tend to seperate, while kinking. Both tilt and kink effects lead to localized current concentrations, where sheetlike structures develop in front of the tilt-displaced fluxropes, locating near-singular peak current values. The kink deformations lead to helically structured current deformations, and ray-traced views of current distributions, as well as quantifications of the electric field component parallel to the field lines, show a rich variety of possible reconnection sites. 

Our 3D setup is complementary to the tilt evolutions known from laboratory plasma spheromak or FRC experiments, where compact toroids with mainly poloidal field are known to be unstable~\citep{iwasawa00}, while sheared toroidal rotation may reduce tilt mode growthrates~\citep{belova00}. In our study, toroidal effects have been eliminated, focusing instead on parallel straight fluxropes that carry opposite currents. We envision our local box simulations to represent conditions likely found in e.g. the highly structured solar corona, where myriads of loops and current channels feature. Indeed, we here propose the tilt-kink instability route as an as yet unexplored pathway for initiating violent coronal mass ejections accompanied by solar flares. The double current channels mimic the top coronal regions of two adjacent flux bundles, with our vertical $z$-axis then running parallel to the solar surface. EUV views on the solar corona typically show many adjacent flux systems, and our setup would connect to an overall quadrupolar current pattern, despite an essentially bipolar magnetic setup (set by our $B_z$ component). The fact that the strength of the fluxrope-aligned $B_z$ component is a decisive factor to stabilization, suggests possible scenarios where its evolution due to radial expansion (i.e. cross-sectional inflation) of fluxropes and flux conservation arguments may render previously stable into unstable setups. The telltale difference with kink or torus instability pathways is the involvement of multiple fluxropes, and their tendency to separate. Recently, detailed analysis of (partially) erupting magnetic flux ropes showed evidence for stable `double-decker' configurations existing for hours before the eruption~\citep{liub12,cheng14}, where double fluxrope dynamics is at play. We here suggest that tilt (or coalescence) type driven interactions may well be involved in the eruption triggers.
The fact that singular 3D current sheets develop as a consequence of the tilt makes these sheets a likely site for particle acceleration and associated flaring events. Follow-up work should study the necessary extensions to models beyond resistive single-fluid MHD, such as modifications due to Hall-effects or in multi-fluid conditions, quantifications of particle acceleration aspects, by studying how the established electric and magnetic field distributions can lead to high energy tails in particle velocity distribution functions. Future studies can also incorporate additional physics and/or more realistic global setups, including curvature and line-tying of the adjacent loop systems, solar coronal radiative losses and anisotropic thermal conduction aspects, or can even make the step to true multi-scale kinetic to fluid simulations treatments (achievable with e.g. the multi-level multi-domain approach~\citep{innocenti13,beck14} or fluid-kinetic particle-in-cell means~\citep{markidis14}).

\begin{figure}
\begin{center}
\FIG{
\resizebox{0.48\textwidth}{!}
%{\includegraphics{figures/synAIA193_30_60_6}}
{\includegraphics{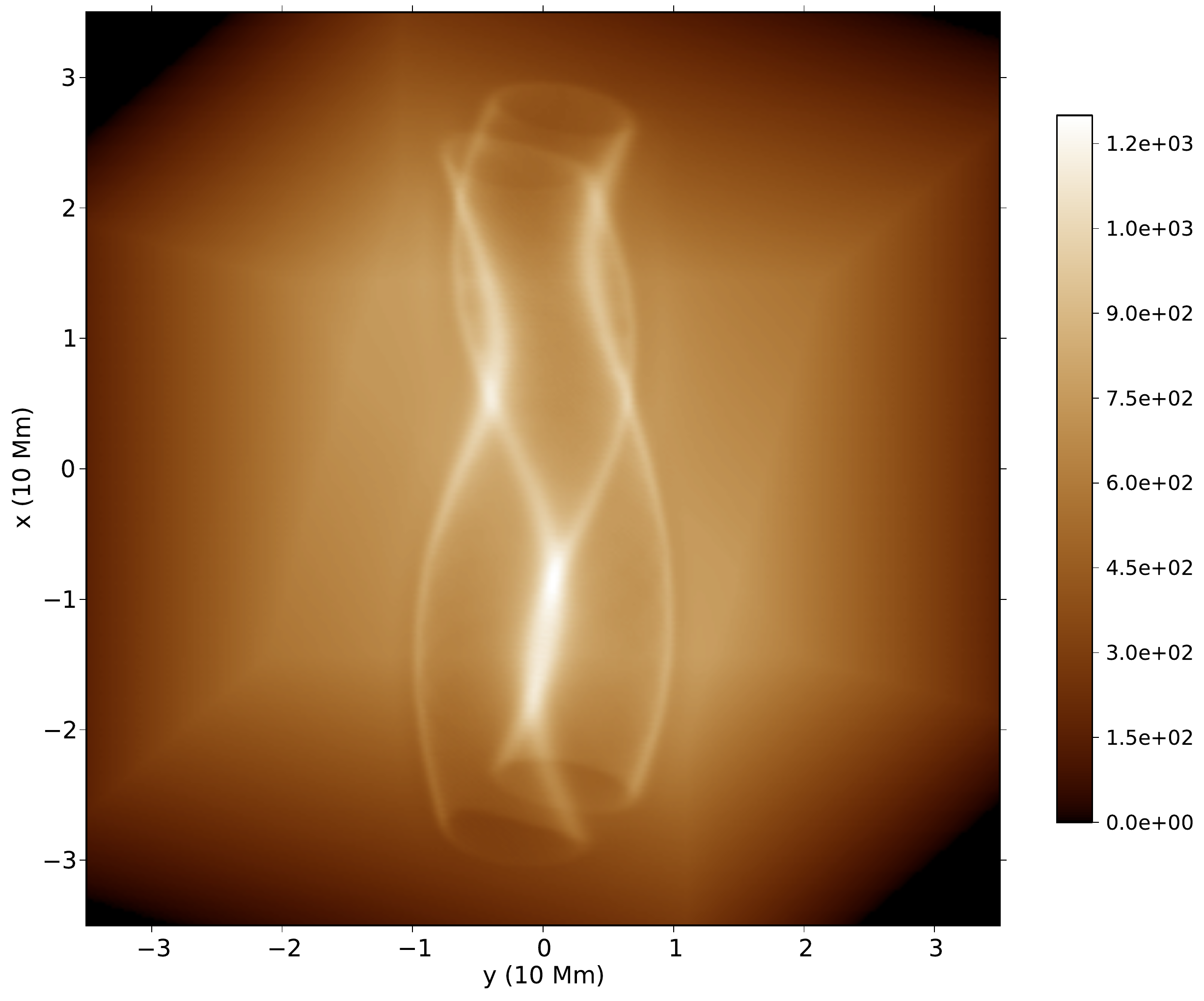}}
\resizebox{0.48\textwidth}{!}
%{\includegraphics{figures/synAIA193_30_60_7}}
{\includegraphics{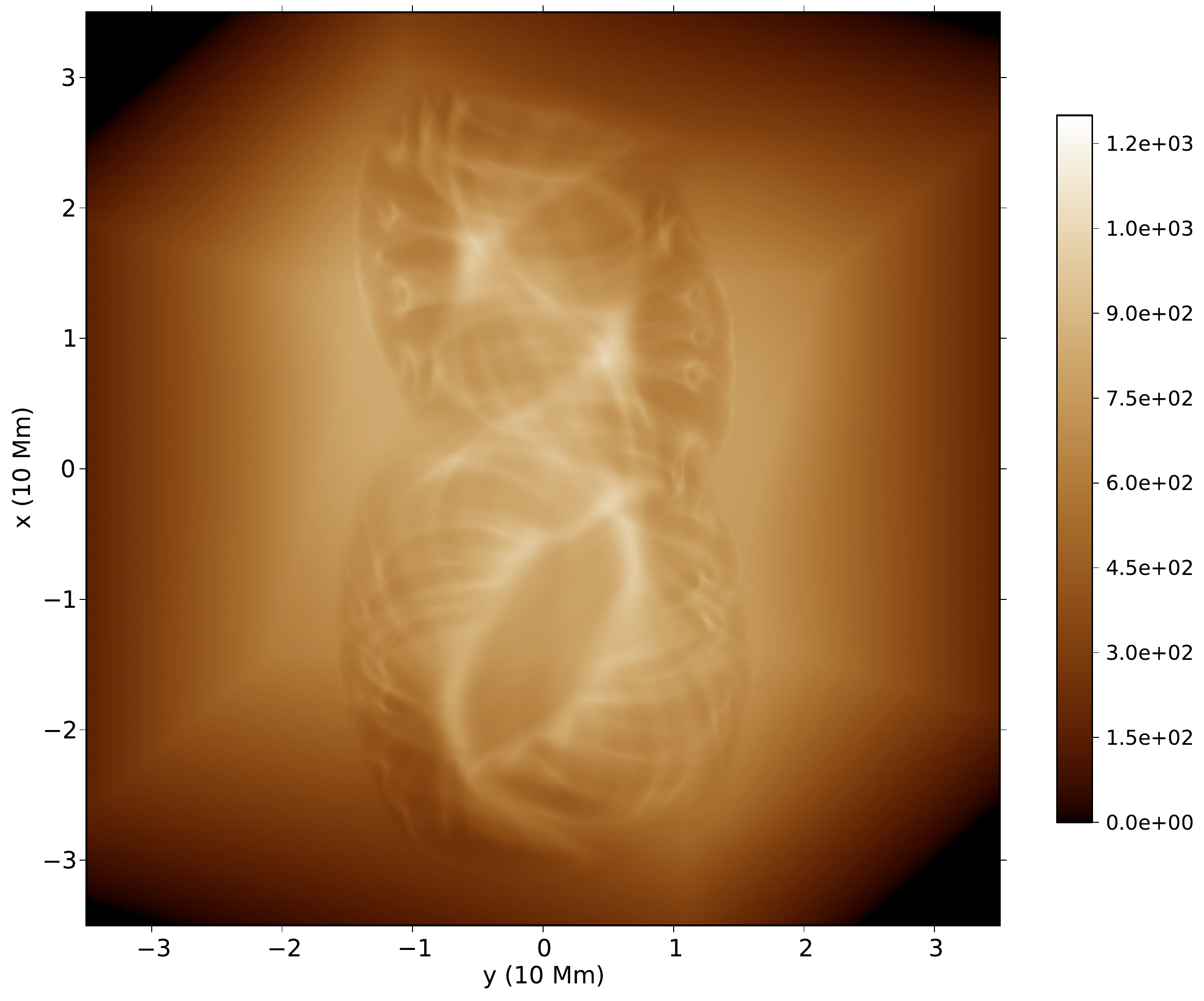}}
}
\end{center}
\caption{Two consecutive synthetic SDO/AIA views at wavelength band 193 \AA, for case DD3d with $B_z=1$, at times $t=6$ and $t=7$. Note how the double current system shows distinct large-scale tilt ($t=6$) evolving to tilt-kink ($t=7$) dynamics, followed by further fragmentation. With the vertical direction interpreted as running parallel to the solar surface, these views show morphological features that would allow to distinguish them from pure kink or torus unstable fluxrope evolutions.}
\label{fig10}
\end{figure}

To demonstrate how the tilt-kink instability would appear in current observations, we use our case DD3d, which has $B_z=1$ and plasma beta around unity (value 0.6 external to both fluxropes), to produce synthetic EUV views typical for the SDO Atmospheric Imaging Assembly (AIA) instrument. To do so, we must restore dimensions, and using a unit of length $L_{\mathrm{u}}=10$ Mm, temperature $T_{\mathrm{u}}=10^6$ K, and number density $n_{\mathrm{u}}=10^9 \, \mathrm{cm}^{-3}$, we attain approximate coronal conditions where the two parallel fluxropes have lengths of 60 Mm, and each has a radius of about 2500 km. A unity dimensionless field strength then translates to $\sqrt{\mu_0 2 n_{\mathrm{u}} k_{\mathrm{B}} T_{\mathrm{u}}}\approx 1.86 \mathrm{G}$, so that the field surrounding the fluxropes has a strength of few Gauss far away, with an initial $t=0$ range which has value 2.63 Gauss far away, reaching up to about 5 Gauss throughout the flux ropes. Although this is certainly on the low side for actual fluxropes or coronal loop complexes, we note that one could exploit more freedom in the equilibrium setup to arrive at lower plasma beta, overdense, hot loops as typical for coronal conditions that still are liable to tilt-kink evolutions. In any case, the temperature-density variation is the essential ingredient needed for generating synthetic views in typical SDO/AIA wavelength bands, which require a line of sight integration of electron number density squared, multiplied with the instrumental response function for the given density-temperature regime at the wavelength of interest. The latter can be obtained from the AIA analysis routines in SolarSoft~\citep{Boe12}. Figure~\ref{fig10} shows two snapshots as viewed in the 193 \AA~wavelength band, at dimensionless times 6 and 7. 
This wavelength band primarily picks up emission from 1.5 MK plasma regions, and for the timespan prior to the $t=6$ frame (since the corresponding time unit is of order 85 seconds, this is about 8.5 minutes), this view shows two parallel hollow cylindrical plasma columns, in accord with the pressure (and temperature) variation needed to realize ideal MHD equilibrium. Suddenly, the kink deformation sets in and this is here seen in the SDO/AIA view as the large-scale undulations of both current channels. About 85 seconds later, the $t=7$ frame shows the helical substructure induced by the kink deformations, as it translates to these density-temperature views. At later times, the fragmentation process is also seen. In future work, we will address in more detail how the different wavelength bands of the SDO/AIA instrument can be used to seek out clear indications of tilt-kink evolutions,
for parameters that match even better with solar coronal values.

\acknowledgments
This research was supported by projects GOA/2015-014 (2014-2018 KU Leuven),
FWO Pegasus, and the Interuniversity Attraction Poles Programme
by the Belgian Science Policy Office (IAP P7/08 CHARM). The
simulations used the VSC (flemish supercomputer center) funded by Hercules
foundation and Flemish government.

%\bibliography{apjsubmission}

\end{document}